\documentstyle[aps,twocolumn,prl,epsfig]{revtex}
\begin{document}
\title{Effect of granularity on the insulator-superconductor transition in ultrathin Bi films}
\author{G. Sambandamurthy, K. Das Gupta, and N. Chandrasekhar}
\address{Department of Physics, Indian Institute of Science, Bangalore 560 012, India}
\maketitle
\begin{abstract}
We have studied the insulator-superconductor transition (IST) by tuning the thickness in quench-condensed $Bi$ films. The resistive transitions of the superconducting films are smooth and can be considered to represent ``homogeneous" films. The observation of an IST very close to the quantum resistance for pairs, $R_{\Box}^N \sim  h/4e^2$ on several substrates supports this idea. The relevant length scales here are the localization length, and the coherence length.  However, at the transition, the localization length is much higher than the superconducting coherence length, contrary to expectation for a ``homogeneous" transition.  This suggests the invalidity of a purely fermionic model for the transition. Furthermore, the current-voltage characteristics of the superconducting films are hysteretic, and show the films to be granular. The relevant energy scales here are the Josephson coupling energy and the charging energy.  However, Josephson coupling energies ($E_J$) and the charging energies ($E_c$) at the IST, they are found to obey the relation $E_J < E_c$. This is again contrary to expectation, for the IST in a granular or inhomogeneous, system. Hence, a purely bosonic picture of the transition is also inconsistent with our observations.  We conclude that the IST observed in our experiments may be either an intermediate case between the fermioinc and bosonic mechanisms, or in a regime of charge and vortex dynamics for which a quantitative analysis has not yet been done.
\end{abstract}

\section{Introduction}

The interplay between disorder and superconductivity in two dimensions has been an active field of study during the last decade. Weak localization in two dimensions~\cite{c1tvr} is a phenomenon where electronic states are localized by any arbitrary amount of disorder in the absence of interaction, resulting in non-metallic behavior.  Superconductivity is an effect in the opposite extreme, in which phase coherence is established due to electron-electron interaction, across the entire length of the sample. The interplay between these two opposing phenomena has led to various interesting results~\cite{c1gold}. 

When quench condensed, the properties of many elements change drastically. Some elements (e.g. $Ga$) show enhanced superconducting transition temperatures, whereas others (e.g. $Bi$) are found to be superconducting only in amorphous form~\cite{c1buchil}. $Ge$ and $Sb$ which are not superconducting in amorphous or crystalline forms, show signs of superconductivity when mixed with materials such as $Au$, $Cu$ and $Ag$.  Signatures of superconducting transitions have been observed in noble metal thin films when deposited at room temperatures on $Ge$ substrates also. Strongin and co-workers first studied quench condensed films grown on thin $Ge$ underlayers~\cite{c1stro1}. A variety of materials and substrates were investigated, the important finding being that films grown on $Ge$, $SiO$ showed measurable conductance even when they were a few monolayers thick, but films on glass, $LiF$ showed measurable conductance only at higher thicknesses. Therefore, films quench condensed on underlayers such as $Ge$ are classified as ``homogeneous" and films grown directly on substrates, ``granular". 

There are some important differences between these ``homogeneous" and ``granular" films, which have been reported~\cite{c1gold}. Granular superconducting films show ``local superconductivity" i.e., a drop in resistance at the bulk transition temperature (bulk $T_c$) value, but develop an upturn at lower temperatures. Thicker films eventually go completely superconducting. This behavior has been explained by the fact that individual grains go superconducting at the bulk $T_c$ and phase coherence is achieved locally but not across the entire sample. Competition between Josephson coupling energy $E_J$ and charging energy $E_c$ is known to be the driving force for zero temperature phase transitions between the superconducting and insulating states in artificial arrays, films and bulk materials. Due to Cooper pairing, there are no free electrons in these systems, and conduction is due to the tunneling of Cooper pairs from one superconducting grain to another.  In such films, the inter-grain capacitance is usually larger than the capacitance of the grain to ground. This charging energy opposes this tunneling, so that the pairs may become localized. This mechanism is referred to as the bosonic mechanism of the suppression of superconductivity.  

On the other hand, homogeneous films show smooth transitions to the zero resistance state. However, in homogeneous films, $T_c$ is greatly suppressed from the bulk value as the films are made thinner.  Conduction in these ultrathin films is completely different from bulk materials and transport mechanisms such as hopping conduction, classical percolation dominate the properties. In these materials, screening is reduced due to the disorder,resulting in a decrease of the attractive interaction required for Cooper pairing. This reduces the transition temperature.  This mechanism is completely different from the previous one, since the key idea here is the complete disappearance of Cooper pairs. 

The insulator-superconductor (I-S) transition has been extensively investigated  over the last decade, in a variety of systems such as thin films~\cite{c1havi,c1beas}, single Josephson junction~\cite{c1pent}, arrays~\cite{c1van} and one dimensional wires~\cite{c1tink}.  The transition can be tuned by changing a parameter such as disorder~\cite{c1havi}, carrier concentration~\cite{c1hebpa} or applied magnetic field~\cite{c1sitb}.  At very low temperatures this transition can be considered a continuous quantum phase transition~\cite{c1shah}. A putative film with a $T_c$ = 0, separates the films showing insulating behavior from those showing superconducting behavior.

\section{Experiment and Observations}

In this paper, we report results on the insulator-superconductor transition (IST), tuned by changing the thickness of quench-condensed $Bi$ films. The resistive transitions of the superconducting films are smooth and can be considered to represent ``homogeneous" films. The IST occurs very close to the quantum resistance for pairs, $R_{\Box}^N \sim h/4e^2$ on several substrates. The IST in homogeneous films can be imagined as the point where the effect on the transport properties, by the localization of electrons and superconducting coherence become comparable~\cite{c1loccoh,c1taop}.  However, at the transition, the localization length is found to be much higher than the superconducting coherence length, contrary to expectation for a ``homogeneous" transition.  This suggests the invalidity of a purely fermionic model for the transition. Furthermore, the current-voltage characteristics of the superconducting films are hysteretic, and show the films to be granular. The relevant energy scales here are the Josephson coupling energy and the charging energy.  However, Josephson coupling energies ($E_J$) and the charging energies ($E_c$) at the IST, are found to obey the relation $E_J < E_c$. This is again contrary to expectation, for the IST in a granular or inhomogeneous, system. Hence, a purely bosonic picture of the transition also appears inconsistent with our observations. 

The experiments were done in a UHV cryostat, custom designed for {\it in-situ} experiments and is described in~\cite{gsmssc}. Pumping is provided by a turbomolecular pump, backed by an oil-free diaphragm pump. A completely hydrocarbon free vacuum $\le 10^{-8}$ Torr can be attained. The substrate is amorphous quartz of size2.5cm X 2.5cm and is mounted on a copper cold finger whose temperature can be maintained down to 1.8K by pumping on the liquid helium bath. The material ($Bi$) is evaporated from a Knudsen-cell with a pyrolytic Boron Nitride crucible, of the type used in Molecular Beam Epitaxy (MBE). $Bi$ is evaporated from the cell at $650^{o}$ C, into a 4-probe resistivity measurement pattern by using a metal mask in front of the substrate. Successive liquid helium and liquid nitrogen cooled jackets surrounding the substrate reduce the heat load on the substrate and provide cryo-pumping. This produces an ultimate pressure $\sim 10^{-10}$ Torr in the system. The metal flux reaching the substrate is controlled using a carefully aligned mechanical shutter in the nitrogen shield. The thickness of the film is increased by small amounts by opening the shutter for a time interval corresponding to the desired increase in thickness. A quartz crystal thickness monitor measures the nominal thickness of the film. Electrical contacts to the film are provided through pre-deposited platinum contact pads ($\sim 50 \AA$ thick). $Ge$ underlayers are deposited on one side of the substrate (a-quartz) before loading the substrate into the cryostat. Separate electrical connections to films on $Ge$ and on bare a-quartz allow us to study both the films simultaneously. I-V's and electrical resistance measurements are done using a standard d.c.current source (Keithley model 220/224) and nanovoltmeter (Keithley model 182) and elctrometer (Keithley model 6514). 

Figure 1. shows the I-S transition in Bi films quench-condensed on 10 $\AA$ of $Ge$ underlayer. Even though there are variations in the value of $R_{\Box}^N$ and transition temperature $T_c$ between films of same thickness quench condensed on different substrates, the value of $R_c$ is found to be close to $h/4e^2$ for $Bi$ films quench condensed on a-quartz, a-quartz with 10 $\AA$ $Ge$ underlayer and films quench condensed on a solid inert layer of $Xe$ in our experiments. The evolution of $R_\Box$ vs T for all the films look similar to the ones shown for $Ge$ underlayer. 

The superconducting transition temperatures of thin films decrease as the thickness $d$ is reduced ($R_{\Box}^N$ is increased). Strong disorder (high $R_{\Box}^N$ or low thickness $d$ with $k_Fl\ll 1$) localizes electron wave functions, increases inelastic scattering rate, suppresses $N(E)$ (the density of states near Fermi energy $E_F$) and ultimately causes a metal-insulator transition. The destructive effect of increasing normal state sheet resistance on superconductivity in 2D has been treated theoretically as a competition between disorder and interaction effects. It has been shown within the frame work of the BCS theory that weak localization of electrons leads to an effective increase in Coulomb repulsion, and corresponding decrease in transition temperature $T_c$. Finkel'shtein~\cite{c3fink} has shown that the lowering of $T_c$ from the bulk value $T_{co}$ follows the relation,\begin{equation}\label{c3fink}\frac{T_c}{T_{co}}=exp\left(-\frac{1}{\gamma}\right)\left[ \frac{\left(1+\frac{(r/2)^{1/2}}{\gamma-r/4}\right)}{\left(1-\frac{(r/2)^{1/2}}{\gamma-r/4}\right)}\right]^{1/\sqrt{2r}}\end{equation}where r is the reduced film sheet resistance ($r = R_{\Box}^Ne^2/2\pi^2\hbar$, measured in units of $\approx 81 k\Omega$) and $\gamma = 1/ln (kT_{co}\tau/2\pi\hbar$) characterises the ratio of the bulk critical temperature $T_{co}$ and the elastic scattering frequency $\tau^{-1}$. $Bi$ in bulk, crystalline form does not superconduct and the ``bulk" critical temperature is normally taken to be the thick film value of 6.10K. The important observation from this equation is that the reduction of $T_c$ from its bulk value does not depend on any intrinsic material property and only depends on $\tau$. 

From this equation we can calculate the reduction in $T_c$ (from the bulk value $T_{co}$) for two types of films, films on $Ge$ underlayers and for films on bare a-quartz. The purpose of showing the data for two different substrates will be discussed later. This is plotted against the reduced r (of equation 3.8) in Fig. 2. The solid lines show the function for different values of $\tau$. From the figure it is clear that except for a small region for the films on bare quartz, the results are not consistent with the Finkel'shtein theory. Finkel'shtein's theory was based on a homogenous two-dimensional disordered system with uniform thickness. If it is believed that the presence of a $Ge$ underlayer facilitates a homogenous film growth, at least these films should have shown  reasonable fit to the theory. But we find that such is not the case.  This can be understood by the fact that the ``homogeneity" of the films might be at length scales smaller than the typical thermal length scale, important for electron - electron interaction ($\sqrt{D\hbar/\pi kT}$), which is of the order of a few hundreds of $\AA$ at these low temperatures~\cite{c1adk1}. For the films on bare quartz, in the small region where Finkel'shtein's theory seems to fit, a value of $-\frac{1}{\gamma}$ = 9 gives a mean free path of $\sim 10\AA$, considering a free electron model.

The I-V characteristics were obtained at 2.25K (which is well below the $T_c$) for all the films. They are plotted in Fig. 3. When current is increased fromthe zero value, the voltage jumps to the normal state value at the critical current ($I_c$). Upon reducing the current from the normal state, the voltage returns to zero not at $I_c$, but at a much lower value $I_{min}$.  These observations are consistent with the I-V characteristics of an underdamped
resistively and capacitively shunted Josephson junction (RCSJ).~\cite{BP, ST, MC, PFC}  In our case, due to the large area of the film, 
the realization is that of a Josephson junction array, with values of $E_J$ and $E_C$ that
are characterized by some distribution, the characteristics of which are determined by
the morphology of the film.  Our analysis of these I-V's in terms of an RCSJ model
has been published.~\cite{gsmssc} This suggests that the film is granular. Further evidence for its granular character is discussed below.

\section{Discussion}

We now discuss these results in the framework of existing models/theories for the IST. Before moving on to this discussion, we present results on the structure and morphology of the films, as inferred from reflection high energy electron diffraction (RHEED). Structure and morphology are important parameters, that can influence the properties of the films, and therefore, conceivably the IST as well. RHEED studies on Bi films, grown on various substrates and underlayers, show that the Bi is almost amorphous.  A RHEED picture is shown in Fig. 4. Based on the Scherrer formula for the peak broadening, we estimate that films thicker than 10 $\AA$ are composed of clusters that vary in size from 25 $\AA$ to 100 $\AA$~\cite{gsmssc}. Since the information obtained is in reciprocal space, it is difficult to comment on the real space surface morphology.  Our RHEED observations are consistent with previous STM work.~\cite{vall} We therefore assign an average size to the clusters of $50 \AA$. This yields a spacing between clusters of approximately $150 \AA$ (considering hemispherical clusters, and using conservation of deposited material) for a film close to the IST, which has a normal state resistance of $6.25 k\Omega$.  It turns out that these parameters, the size of a grain or island, and the average distance between islands are important parameters, a fact that is obvious for the bosonic mechanism.

Our observations suggest that the film can be considered as a random  array of Josephson junctions, which are shunted by a resistance. Consequently the  resistively and capacitively shunted junction (RCSJ) model~\cite{BP} can be used to describe the hysteretic behaviour of the I-V curves, with the capacitance being the intrinsic capacitance of the junction. From the ratio of $I_{min}/I_c$, the value of the admittance ratio ($\beta$) can be calculated ~\cite{ST,MC,PFC}.  Here $\beta = \omega_{c}C/G$, where $\omega_{c}$ is ($2e/\hbar)I_c R_s$.  C is the intergranular capacitance and G the normal state conductance of the array. We wish to point out that these are lumped parameters, which characterize the whole array. From the values of $\beta$, the intergrain capacitance is calculated. The charging energy $E_c (= e^2/2C)$ and the Josephson coupling energy $E_J = \hbar I_c/2e$ are calculated for all the film thicknesses studied. These values are calculated using single value of C and G, which correspond to capacitance and conductance of the array. C and G will have a range of values, the distribution of these values and the moments of the distribution will of course depend on film thickness. We measure the critical current $I_c$ for different films at T = 2.0K (lower than $T_c$ for all the thicknesses studied) and calculate the relevant energy parameters such as the charging energy ($E_c$), Josephson coupling energy ($E_J$) etc. Fig. 5. shows the ratio of the Josephson coupling energy to the charging energy vs the sheet resistance for the films quench condensed on $Ge$ underlayer at the temperature where the I-V's were acquired, T=2.25 K. We find that even though the IST occurs near $R_c$, the relevant energy scales become equal at a much higher thickness. This suggests that the a purely bosonic mechanism may be an incorrect picture for understanding the destruction of superconductivity in these films. We next investigate the validity of the fermionic mechanism.

To check whether the fermionic mechanism is a good representation of the physical mechanism, we estimate the electron localization length from the high temperature resistance data.  In strongly disordered films, the temperature dependence of the resistivity is of the form $R = R_0 exp[(T_o/T)^{\alpha}]$, where $\alpha$ varies from 0.75 for collective variable range hopping, 0.5 for hopping dominated by Coulomb interactions (Efros-Shklovskii correlated hopping) and 0.33 for Mott variable range hopping. In the weak-localization regime, the conductivity shows a logarithmic dependence on temperature,\begin{equation}\sigma = \sigma_0 + e^2p lnT/(\pi h)\end{equation} where p is a coefficient determined by the scattering mechanism for electrons. In estimating the localization length, we neglect interactions. In previous work, which involved studies of quench condensed films on different substrates whose dielectric constants varied from 1.5 (for solid Xe underlayers) to 15 (for Ge underlayers), we have demonstrated that the IST is robust and unaffected by the dielectric constant of the substrate~\cite{prb1}. This is our justification for neglect of interactions. We use the theory of W\"olfle and Vollhardt~\cite{wv} which describes the transition from weak to strong localization, neglecting interaction. Their result is \begin{equation}\hbar/(e^2R_{\Box}) = 1/(2\pi^2) [ln(1+y^2)](1+y)exp(-y)\end{equation}where $y = L/\xi_{loc}$. $\xi_{loc}$ is a localization length, related to the elastic mean free path l by $\xi_{loc} = l exp(\pi k_F l/2)$, where $k_F$ is theFermi wave vector. Here the sample size is regarded as a cutoff length due to inelastic scattering $L = D T^{-p}$, where D is a diffusion constant for electrons. Knowing the resistivity at a suitably high temperature (where there is no observable temperature dependence) and its variation with temperature (at lower temperatures), the various parameters can be determined. 

We determine the superconducting coherence length from upper critical field data, which has been presented in a separate publication~\cite{prbrap}. We have determined $B_{c2}(T)$ of our films, from the resistive transition in a perpendicular magnetic field. The convention that we have followed is to define $B_{c2}$ as the field at which the sheet resistance is half its normal state value, $R_{\Box}^N$. We then use the Ginzburg-Landau definition of $B_{c2}(T) = \Phi_0/(2\pi\xi^2)$, to determine the coherence length. Fig. 6 shows the variation of the superconducting coherence length and the localization length with sheet resistance of the films. We note here that the coherence lengths of the thinnest films which show a decreasing resistance with temperature, cannot be measured experimentally, since the lowest temperatures accessible to us in our apparatus is 1.6 K. We therefore extrapolate the coherence lengths to higher sheet resistances. From the behavior of $\xi$ at lower sheet resistances, this approximation is clearly justified.  As is evident from Fig. 6, the IST occurs at a point where the localization length is much larger than the coherence length, $\xi_{loc}$ is $800 \AA$, whereas $\xi$ is only $25 \AA$. The ratio $\xi_{loc}/\xi$ is 32 in our study of quench condensed Bi films. This is to be contrasted with the results of Kagawa {\it et al.}~\cite{c1loccoh}, who found a ratio of two for Pb films. Whether this difference is due to the different materials studied, or the differing deposition geometry, is unclear to us at the present time.

Other models for the destruction of superconductivity in such granular systems have been considered for proximity arrays of varying geometry (ratio of the separation of the superconducting regions to their size).~\cite{lark1,lark2} Although these papers consider a mechanism that appears to be intermediate between the bosonic and fermionic mechanisms discussed above, there are several constraints on the sample geometry and physical properties. The ratio of the spacing between islands $b$, to the island size $d$ should obey the relation $ln(b/d) \ge 3$.  In our films, this condition is clearly not met, since we have $b/d \sim 3$, and $ln(b/d) \sim 1$.  Further, the authors consider a case where there is a fairly large tunneling conductance between the substrate and each superconducting island. This large tunneling conductance can help couple the islands, so that the conventional Coulomb blockade effect is suppressed. This is the physical reason for a mechanism for the IST that is not purely bosonic. The tunneling conductance between the islands and the substrate are expected to be quite different for Bi films on Ge and a-quartz, since the dielectric constants of the two materials differ by a factor of three. In Fig. 1, we show fits to Finkel'shtein's theory for these two substrates. We reiterate the point that the fits are reasonable only over a small region, for the quartz substrate. Since we expect the tunneling conductances to differ, we expect to be in a regime where the model of Feigel'man {\it et al.} may be relevant.The large tunneling conductances between the substrate and thesuperconducting islands results in disorder enhanced multiple Andreev reflection. In our studies, we have used several substrates, ranging from solid xenon to Ge, but we still observe a robust IST close to a thickness of $25 \AA$ and a $R_{\Box}$ of $h/4e^2$~\cite{prb1}. Over this range of dielectric constants, it is natural to expect the island-substrate (or underlayer) tunneling conductance to vary substantially. However the robust nature of the IST suggests that Andreev reflection and the associated physics is not relevant in our case.

Considerations of phase fluctations in arrays of regular as well as random Josephson junctions have shown that both vortex-unbinding transition, as well as the charge unbinding transition can occur~\cite{fazio}.  Both of these transitions can occur, depending on the ratio between the charging energy and the Josephson coupling energy.  Fazio and Schon have set a limit for $E_J/E_c$ of $2/\pi^2$ for the boundary between the insulator and the superconductor.  In our case, the transition occurs at a small $E_J/E_c \simeq 10^{-3}$. At such small values, Cooper pairs are expected to remain frozen (no long range phase order), but the single electron dynamics is the same as in the normal state, resulting in an insulating state~\cite{fazio}. In our experiments, we are in a regime where the normal state conductance is close to 1, the films are granular, and $E_J/E_c$ is very small. This is a regime, for which a quantitative analysis has not yet been done. Since an experimental realization of this limit has now been observed, we hope that this work would stimulate such an analysis. In this regime, where the normal state conductance in high, single electron processes other than Andreev reflection, may be important.

\section{Conclusions}

In conclusion, we find an IST very close to the quantum resistance for pairs, $R_{\Box}^N \sim h/4e^2$ for Bi films on several substrates, however, at the transition, the localization length is much higher than the superconducting coherence length. This is contrary to expectation for the IST in a ``homogeneous" film.  
Therefore, we explore models other than a purely fermionic model for the transition. 
The current-voltage characteristics of the superconducting films are hysteretic, and can be fitted to an RCSJ model, suggesting that the films are granular. In this case, the relevant energy scales here are the Josephson coupling energy ($E_J$) and the charging energy ($E_c$). However, at the IST, we find that $E_J < E_c$. 
This is in conflict with the conventional model for the IST in a granular or inhomogeneous system. A simple bosonic picture of the transition is also inconsistent with the reported observations.  In the experiments reported here, the dimensionless conductance is nearly unity, and the ratio of Josephson energy to charging energy very small. This is a regime which has not been investiagated theoretically, due to the difficult nature of the analysis. Our observations suggest that such a regime merits further investigation, since this is the most likely experimental realization not only in our study, but in several earlier studies as well.~\cite{c1gold, c1havi, c1hebpa}

{\bf Acknowledgemnets}
This work was supported by DST, Government of India. KDG thanks CSIR, New Delhi for the financial support through a Senior Research Fellowship.  

\newpage
{\centerline {FIGURE CAPTIONS}}

(1) Fig. 1. Insulator - Superconductor transition for a set of $Bi$ films quench condensed at 15K on quartz substrates pre-deposited with 15 $\AA Ge$ underlayer.\\

(2) Fig. 2. $T_c/T_{co}$ for films on $Ge$ and a-quartz, plotted against the reduced resistance. The solid lines show the perdictions of Finkel'shtein's theory for different values of $\tau$. See text for details.\\

(3) Fig. 3. Set of I-V curves for superconducting $Bi$ films which shows hysteretic behavior.  The number beside each I-V is the film thickness in $\AA$.\\

(4) Fig. 4. The RHEED picture from a typical superconducting Bi film during growth. At least one diffuse ring is clearly visible.

(5) Fig. 5. The variation of the ratio of the Josephson coupling energy to the charging energy ($E_J/E_c$) with the normal state sheet resistance $R_N$ for the films in Fig.1. The arrow indicates $R_N = h/4e^2$.\\

(6) Fig. 6. Variation of the localization lengths and coherence lengths with the normal state sheet resistance $R_N$ for the films in Fig.1. The arrow indicates $R_N = h/4e^2$.\\

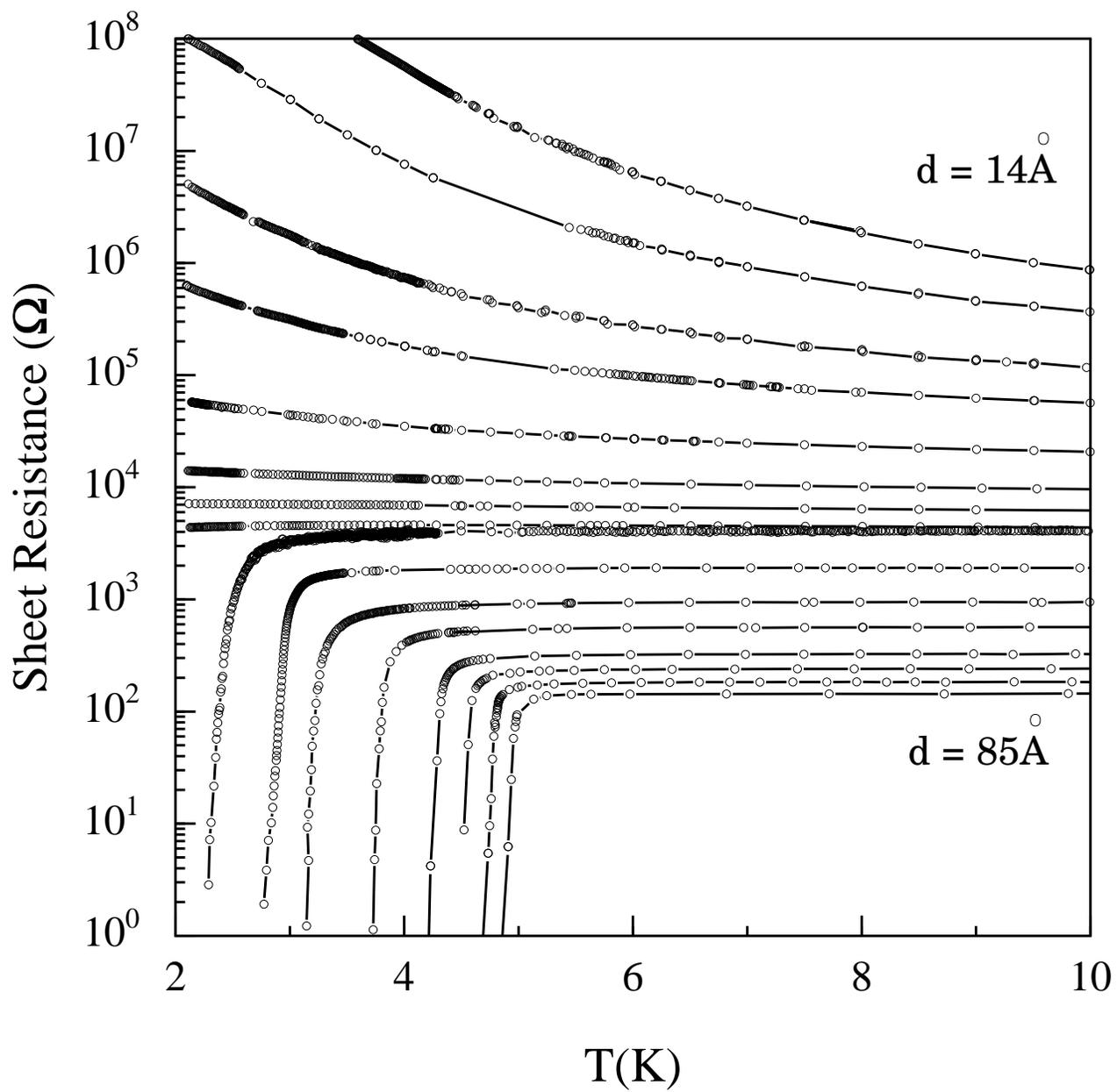
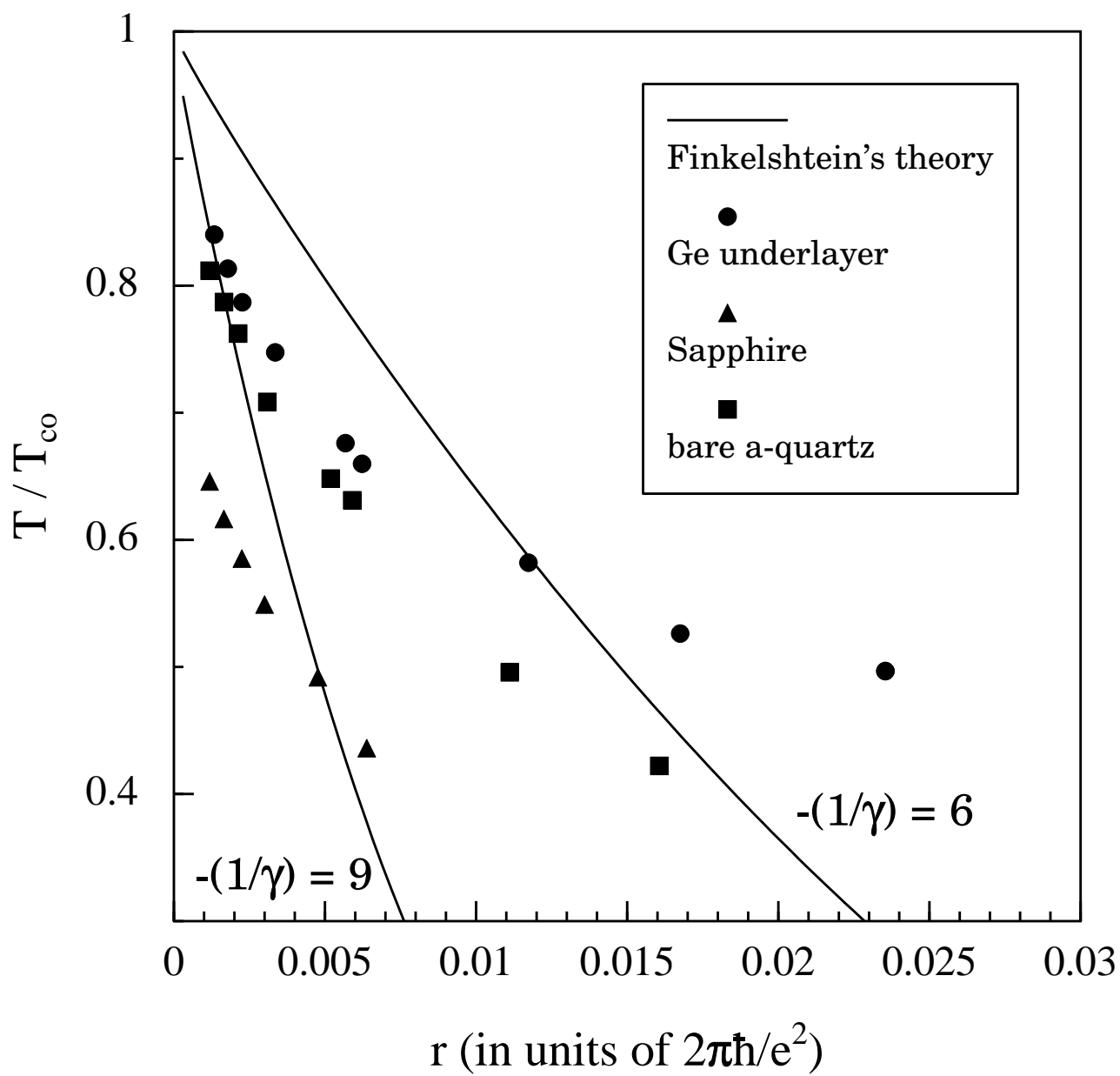
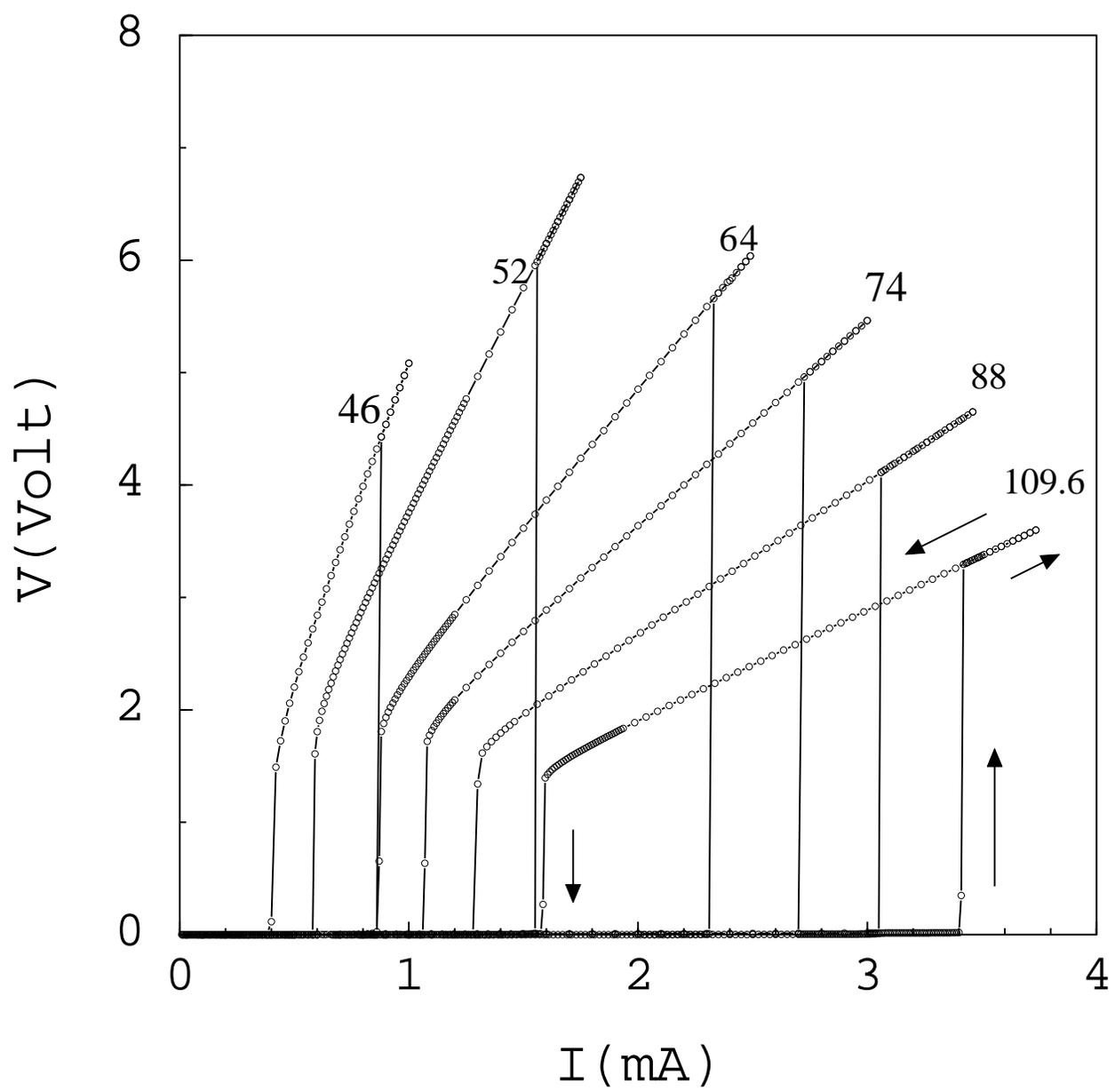
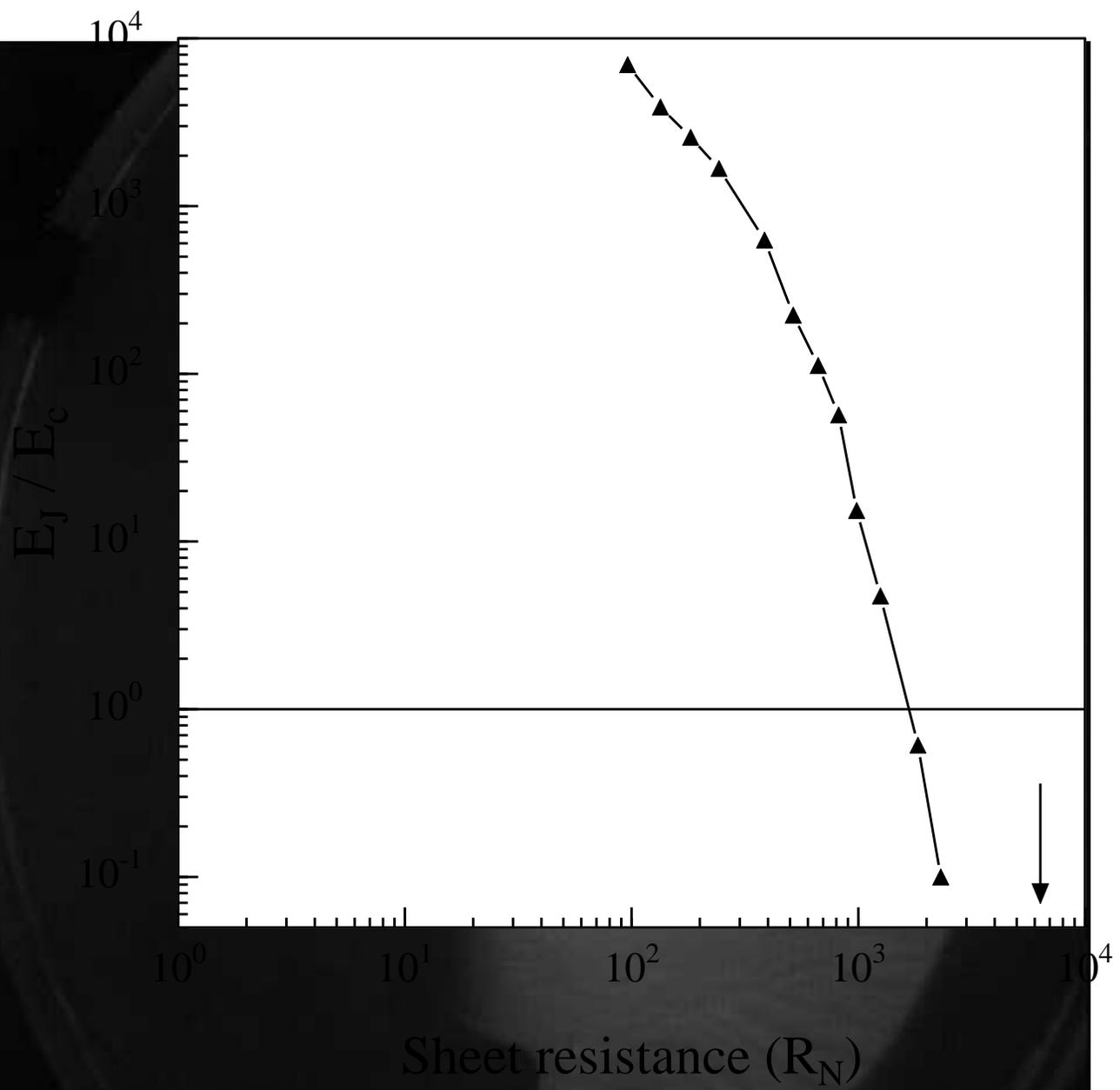
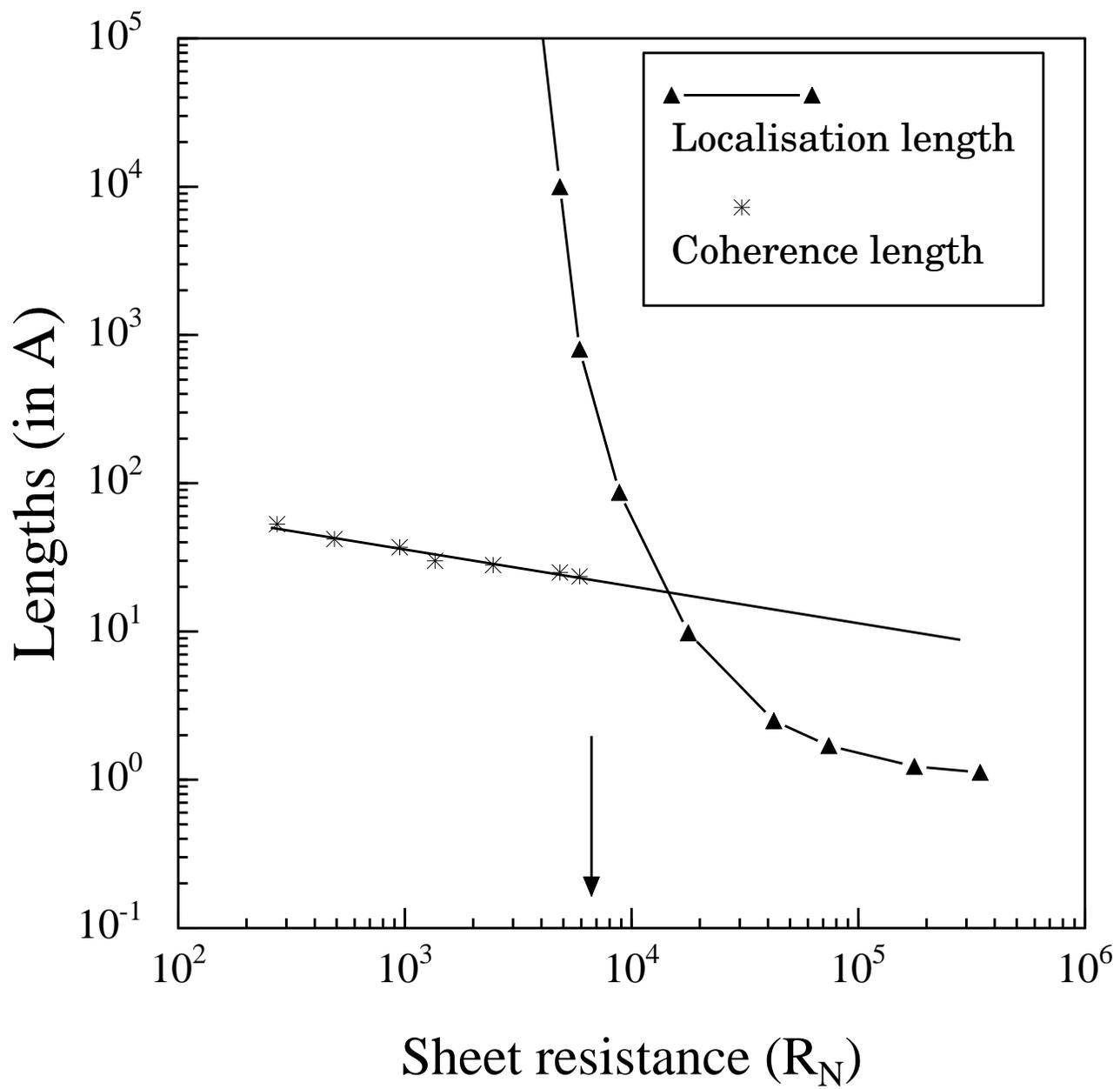
\end{document}